# The DECam Ecliptic Exploration Project (DEEP). VII. The Strengths of Three Superfast Rotating Main-belt Asteroids from a Preliminary Search of DEEP Data

Ryder Strauss[1], Andrew McNeill[1,2], David E. Trilling[1], Francisco Valdes[3], Pedro H. Bernardinelli[4], Cesar Fuentes[5], David W. Gerdes[6,7], Matthew J. Holman[8], Mario Jurić[4], Hsing Wen Lin (林省文)[6], Larissa Markwardt[6], Michael Mommert[9], Kevin J. Napier[6], William J. Oldroyd[1], Matthew J. Payne[8], Andrew S. Rivkin[10], Hilke E. Schlichting[11], Scott S. Sheppard[12], Hayden Smotherman[4], Chadwick A. Trujillo[1], Fred C. Adams[6,7], and Colin Orion Chandler[1,4,13]

[1] Department of Astronomy and Planetary Science, Northern Arizona University, PO Box 6010, Flagstaff, AZ 86011, USA
[2] Department of Physics and Astronomy, Bowling Green State University, Bowling Green, OH, 43403, USA
[3] National Optical-Infrared Astronomy Research Laboratory, Tucson, AZ 85726, USA
[4] DiRAC Institute and the Department of Astronomy, University of Washington, Seattle, USA
[5] Departamento de Astronomía, Universidad de Chile, Camino del Observatorio 1515, Las Condes, Santiago, Chile
[6] Department of Physics, University of Michigan, Ann Arbor, MI 48109, USA
[7] Department of Astronomy, University of Michigan, Ann Arbor, MI 48109, USA
[8] Harvard-Smithsonian Center for Astrophysics, 60 Garden Street, MS 51, Cambridge, MA 02138, USA
[9] Stuttgart University of Applied Sciences, Stuttgart, Germany
[10] Applied Physics Lab, Johns Hopkins University, 11100 Johns Hopkins Road, Laurel, Maryland 20723, USA
[11] Department of Earth, Planetary and Space Sciences, University of California Los Angeles, 595 Charles E. Young Dr. East, Los Angeles, CA 90095, USA
[12] Earth and Planets Laboratory, Carnegie Institution for Science, Washington, DC 20015, USA
[13] LSST Interdisciplinary Network for Collaboration and Computing, 933 N. Cherry Avenue, Tucson AZ 85721, USA

## Abstract

Superfast rotators (SFRs) are small solar system objects that rotate faster than generally possible for a cohesionless rubble pile. Their rotational characteristics allow us to make inferences about their interior structure and composition. Here, we present the methods and results from a preliminary search for SFRs in the DECam Ecliptic Exploration Project (DEEP) data set. We find three SFRs from a sample of 686 main-belt asteroids, implying an occurrence rate of $0.4^{+0.1}_{-0.3}\%$—a higher incidence rate than has been measured by previous studies. We suggest that this high occurrence rate is due to the small sub-kilometer size regime to which DEEP has access: the objects searched here were as small as ∼500 m. We compute the minimum required cohesive strength for each of these SFRs and discuss the implications of these strengths in the context of likely evolution mechanisms. We find that all three of these SFRs require strengths that are more than that of weak regolith but consistent with many cohesive asteroid strengths reported in the literature. Across the full DEEP data set, we have identified ∼70,000 Main-Belt Asteroids and expect ∼300 SFRs—a result that will be assessed in a future paper.

*Unified Astronomy Thesaurus concepts:* Main belt asteroids (2036); Asteroid rotation (2211); Period search (1955); Lomb-Scargle periodogram (1959); Astronomical object identification (87); Astrometry (80)

## 1. Introduction

The small objects of the solar system have an enormous potential for providing insight into its formation and evolution. Among the most well-studied small solar system bodies are the asteroids within the main belt, a region between the orbits of Mars and Jupiter, with heliocentric distances spanning roughly 2–3.5 au. These main-belt asteroids (MBAs) are powerful tools for understanding solar system dynamics for a few key reasons. Their relatively low masses mean that they can often be treated as test particles that trace the dynamical history of the major planets. Because MBAs are so numerous and, compared to other populations like those in the Kuiper Belt, so nearby and bright, they are relatively easy to measure at a very high volume. Thus they provide an enormous statistical constraining power for bulk population characteristics as well as identifying unusual individual objects. As of 2024 April, almost 1.3 million MBAs have been discovered as reported by the International Astronomical Union's Minor Planet Center.[14]

Despite the enormous sample of known MBAs, our understanding of their physical characteristics is subject to several limitations. Fewer than 10 MBAs have been visited by spacecraft, while the vast majority have only been measured from a distance via ground- or space-based telescopes using photometry, spectroscopy, spectrophotometry, resolved imaging, radar, thermophysical modeling, or other remote measurement techniques (J. Veverka et al. 1996; P. C. Thomas et al. 2000; Ž. Ivezić et al. 2001; T. C. Duxbury et al. 2004; F. E. DeMeo & B. Carry 2013; M. Taylor et al. 2016; A. V. Sergeyev & B. Carry 2021). This limits much of our direct knowledge of most MBAs to surface characteristics, with their interiors remaining a mystery.

Fortunately, several techniques provide indirect probes of the MBA interiors. One such approach is to investigate the rotational properties of these objects. As an aspherical solar system object rotates in space, its projected cross-sectional area changes over time. For an unresolved object, this appears to an observer as a periodic change in photometric brightness. The shape of this lightcurve can encode information about the asteroid's rotation period, its shape, and even the orientation of its spin pole, given a sufficient number of measurements (J. Durech et al. 2009; J. Hanuš et al. 2011).

---

[14] https://www.minorplanetcenter.net/

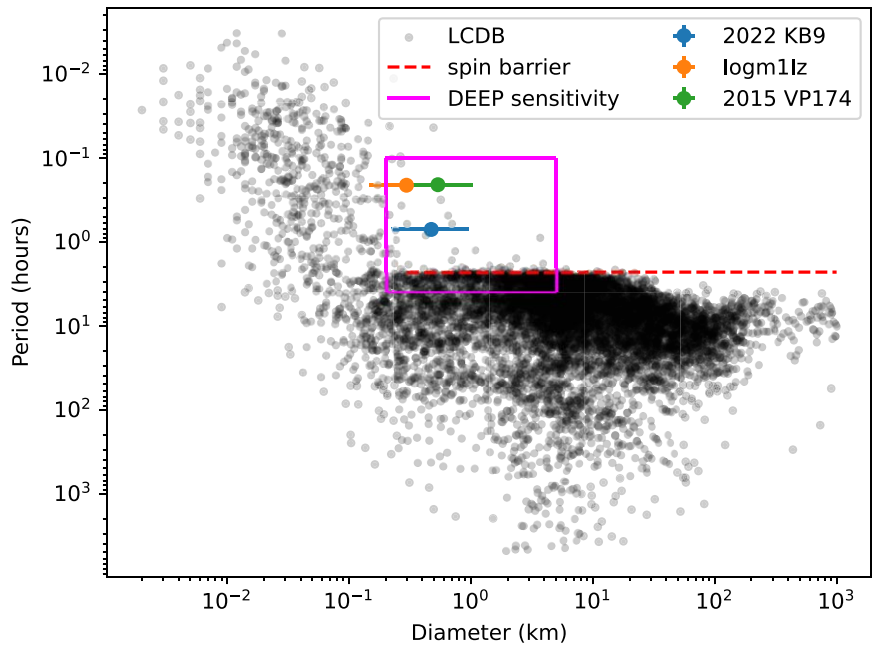

**Figure 1.** Diameter vs. Period for ∼22,000 main-belt asteroids as compiled by the asteroid lightcurve database. The red dashed line at 2.2 hr indicates the spin barrier. Very few asteroids with diameters >0.2 km have rotation periods less than 2.2 hr, making this our "region of interest" for SFRs. Overplotted in magenta is the approximate region to which the DEEP survey is sensitive (The bright limit is the diameter that begins to saturate the Dark Energy Camera (DECam) at 120 s). This region has considerable overlap with the SFR region of interest. The three colored dots indicate the position of three DEEP SFRs presented in this paper. We estimate the diameter of the DEEP objects from their absolute magnitude, assuming a geometric albedo of 0.15. Our estimated diameter uncertainties are based on imposing lower and upper albedo limits of 0.05 and 0.5.

The Asteroid Light Curve Database (LCDB; B. D. Warner et al. 2009) provides an archive of ∼22,000 MBAs with measured rotation periods and lightcurve amplitudes as of 2023 October 1. Figure 1 shows the relationship between asteroid diameter and rotation period for the subset of MBAs with reported LCDB lightcurves. Inspection of this figure reveals an interesting feature: for asteroids with diameter $D > 0.2$ km, almost none have rotation periods shorter than 2.2 hr. This empirical boundary, often referred to as the "spin barrier," demarcates the spin rate above which an asteroid with no cohesive forces aside from self gravity (i.e., a "rubble pile") would be expected to break apart under centrifugal acceleration. For a typical S-type MBA density of roughly 2500 kg m$^{-3}$ (B. Carry 2012), this empirically measured boundary corresponds to the theoretical spin period at which the speed of a particle on the surface of a spherical asteroid is equal to the surface escape velocity.

Figure 1 shows that not all objects have spin periods slower than 2.2 hr—many have spin rates far faster than that. At the smallest sizes, this is straightforward to explain: below diameter $D \sim 0.2$ km, most asteroids are thought to be solid monolithic fragments of rock that can have immense cohesive strength and have nearly arbitrary spin periods without losing material (K. J. Walsh 2018). More compelling are the few fast-rotating MBAs with diameter $D > 0.2$ km. Most MBAs within this size regime are rubble piles—self-gravitating aggregates of smaller pebbles, boulders, and other material. Asteroids that fall within this rubble-pile size regime and rotate faster than the spin barrier are called superfast rotators (SFRs). These objects are powerful tools for understanding the interior of asteroids in their size regime: because they have rotation periods which should not be possible under self gravity alone, their existence requires some considerable internal cohesive strength preventing their breakup (A. McNeill et al. 2018b).

The SFR region of parameter space is challenging to measure due to the temporal resolution required to detect very short lightcurve periods. Many MBA surveys have a 10,000 hr time baseline (decades to years), but the time between individual measurements is also very long (days to months), making it difficult to detect short-term variations (A. N. Heinze et al. 2018; E. C. Bellm et al. 2019; D. E. Trilling et al. 2023; D. Kramer et al. 2023). Those surveys that do image large fractions of the sky at a very fast cadence often do so to a very bright magnitude limit (N. M. Law et al. 2015). While many SFRs have been successfully identified and characterized in the main belt (E. Rondón et al. 2020; F. Monteiro et al. 2020; J. Licandro et al. 2023), a comprehensive study of main-belt SFRs requires a survey that can successfully address each of these limitations.

The DECam Ecliptic Exploration Project (DEEP) is a wide-field solar system survey, primarily designed to discover and characterize objects in the trans-Neptunian region of the solar system (C. A. Trujillo et al. 2024; D. E. Trilling et al. 2024; H. Smotherman et al. 2024; K. J. Napier et al. 2024; P. H. Bernardinelli et al. 2024; R. Strauss et al. 2024). Because it is a survey of the ecliptic plane, however, it is also well suited for the detection of main-belt asteroids. The survey uses the DECam, a high-sensitivity CCD mosaic imager with a three-square degree field of view mounted on the Victor M. Blanco 4 meter telescope at Cerro Tololo Inter-American Observatory in Chile (B. Flaugher et al. 2015). The survey cadence comprises a series of 4 hr long stares, each of which consists of roughly 100 2 minute exposures at a 2.3 minute cadence. Thanks to the large telescope aperture and the sensitivity of DECam, the average single-exposure depth for the DEEP survey is ∼23.5 in the VR filter.

Over the entire survey, DEEP has covered more than 120 unique square degrees of sky to this approximate depth, permitting the detection of subkilometer MBAs at a high volume. The fast cadence of the survey also makes it very effective for identifying short periods: with our 2.3 minute observing cadence, lightcurve periods as short as 8 minutes can be identified with ease. Because of the four-hour time baseline of the DEEP long stares, we are limited to detecting periods shorter than 4 hr, but this is well more than the rotation periods of SFRs. Figure 1 demonstrates DEEP's region of sensitivity within the period-diameter parameter space, particularly its overlap with the SFR region of interest. In addition to its access to very fast-rotating objects, DEEP is sensitive to subkilometer MBAs, the size regime thought to be dominated by rubble piles. Thanks to these factors, DEEP is extremely well suited for detecting superfast rotators in the main belt.

In this work, we present the methods and results from a preliminary search for SFRs in the DEEP main-belt sample. In Section 2 we discuss the methods used for identifying moving objects, measuring their rotation periods, and modeling their cohesive strengths. In Section 3 we present the results from the period analysis. Section 4 provides a discussion of the cohesive strength analysis performed with these lightcurves. In Section 5 we provide a discussion of these results in the context of the formation and evolution history of the solar system, and in Section 6 we summarize this work and our conclusions, and discuss recommended future work.

## 2. Data Acquisition and Processing

Observations were obtained using DECam, a three-square-degree wide-field imager mounted on the Blanco 4 m telescope at Cerro Tololo Inter-American Observatory. Most of the observations are "long stares" during which we observe a single patch of sky with roughly 100 sequential 2 minute exposures. Details about the observation strategy for the DEEP survey are provided in C. A. Trujillo et al. (2024). After observations were complete, the data were reduced and moving objects were identified as part of the DECam Community Pipeline: transient source catalogs were generated from difference images, and trios of sources with a consistent brightness and rate of motion were identified as moving objects (F. Valdes et al. 2014). Across the first 4 yr of the survey, ∼60,000 individual moving objects were identified, ∼18,000 of which were successfully linked to the orbits of known objects listed in the Minor Planet Center database. For each object, we have a list of observations with astrometry and photometry over a single ∼4 hr DEEP long stare.

As a pilot study, this work addresses a subset of these moving objects that were searched for short-period photometric variation. We searched all objects identified from one DEEP observing run on the night of 2022 May 25, covering one individual DEEP field. This observing night was chosen as it corresponded closely to dates during which we were able to obtain observing time on the Lowell Discovery Telescope. To perform this search, we made use of the Lomb–Scargle periodogram, a signal processing technique that produces a power spectrum across a range of periods (or frequencies) within time-series data (N. R. Lomb 1976; J. D. Scargle 1989; J. T. VanderPlas 2018). Any dominant peak within the power spectrum is considered to indicate a strong periodic signal within the data. We applied this method to photometry from 686 objects that were identified within a single night's long stare of DEEP data. To identify peaks in the power spectrum, we implement a peak-finding algorithm that identifies individual local maxima and phases the lightcurve to each of these local maxima. For each period that corresponds to a normalized power greater than 0.5, we visually inspect the phased lightcurve to identify the most likely true period. From this analysis, we can successfully differentiate among objects with no clear rotation period within the 4 hr observing window, objects with a strong periodic signal longer than 2.2 hr but less than 4 hr, and objects with power spectrum peaks at periods shorter than 2.2 hr. This work only addresses the latter of these three; lightcurve analysis of the intermediate-period group will be presented in future work.

While DEEP is excellent at discovering SFRs, it has poor capacity to recover them, except in the unlikely case that an SFR appears in more than one DEEP long stare pointing. With only a 4 hr time baseline informing the orbital arc, the positional uncertainty for a newly discovered MBA will exceed 30′ within roughly a week. To permit the recovery of candidate SFRs from this pilot study, we were allocated observing time on the Lowell Discovery Telescope (LDT) shortly after a DEEP observing run in 2022 May. Thanks to the fast turnaround of the DECam Community pipeline, we had a list of SFRs from our single night of DEEP data identified and prepared for re-observation within three days of their initial measurement. Because it was not yet known whether these objects were known objects with well-characterized orbits, this fast follow up permitted us a unique opportunity to significantly extend the orbital arc and verify our computed rotation periods of DEEP SFRs.

The LDT data were obtained during the nights of 2022 June 3–4 with the VR filter (500–750 nm) on the 11′ Large Monolithic Imager. The images were searched by eye for moving sources that were then successfully linked to the DECam observations using Project Pluto's Find_Orb software.[15] Instrumental magnitudes were obtained using aperture photometry with the Photutils software package (L. Bradley et al. 2024) and were calibrated to a zero-point magnitude computed from Gaia DR3 field stars (Gaia Collaboration et al. 2023).

## 3. Results

Out of the 686 MBAs identified during a single night of DEEP observing, three were found to have rotation periods far shorter than that of the 2.2 hr spin-barrier. This implies an incidence rate of roughly 0.4%. The raw lightcurves, Lomb–Scargle power spectra, and phased rotational lightcurves for each object are shown in Figure 2. For each object, a single strong peak is evident within the power spectrum. When the lightcurve is phased to the period associated with the power spectrum peak, the shape of the lightcurve becomes apparent. These SFRs have respective rotational periods of 0.21 hr, 0.71 hr, and 0.21 hr. The rotational characteristics of these three asteroids are summarized in Table 1.

Because it is advantageous to consider the rotational properties in terms of the approximate size of the object, we compute and report an absolute H magnitude for each object.

---

[15] https://www.projectpluto.com/fo.htm

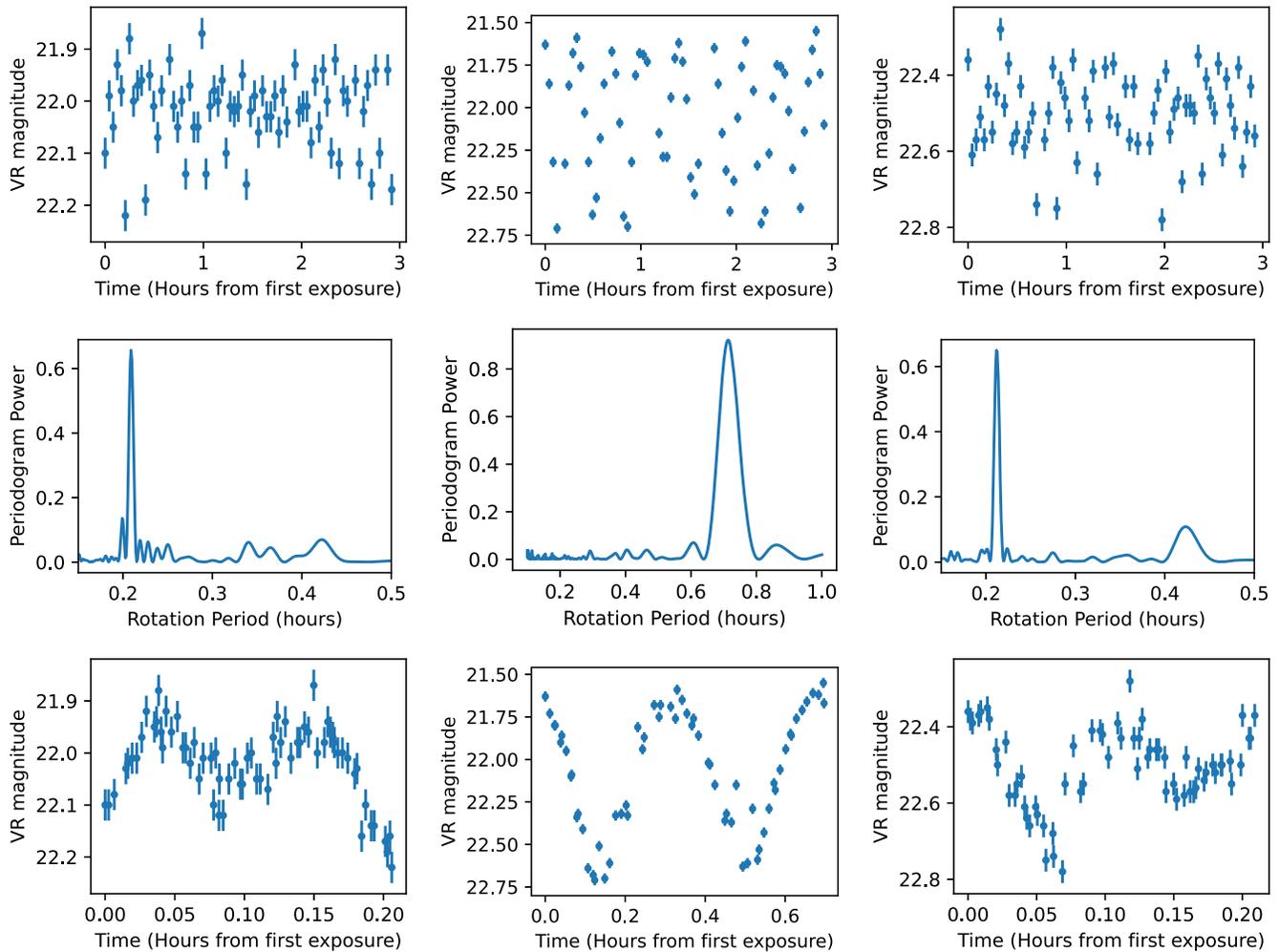

**Figure 2.** Top: unphased lightcurves for the 3 SFRs presented in this work. The scale of the photometric variation is far greater than that of the photometric uncertainties, implying rapid intrinsic variation. Middle: relative power spectra from the periodogram analysis. The periodogram power is presented here as a function of time rather than frequency, and the x-axis has been doubled to account for the double-peaked lightcurve from a full rotation. The strong peaks correspond to the strongest periodic signal within the lightcurve, which we interpret as the most likely rotation period of the object. Bottom: lightcurves phased to the best-fit rotation period for the object.

**Table 1**
Summary of the Rotational Characteristics Derived in this Work

| Asteroid Name | Derived $H$ (Mag) | Rotation Period (hr) | Lightcurve Amplitude (mag) | Approximate Diameter (km) |
|---|---|---|---|---|
| 2022 KB9 | 19.0 | 0.21 ± 0.01 | 0.3 ± 0.1 | $0.5^{+0.5}_{-0.2}$ |
| 2015 VP174 | 19.3 | 0.71 ± 0.01 | 1.0 ± 0.1 | $0.4^{+0.5}_{-0.25}$ |
| Iogm1Zl[a] | 20.3 | 0.21 ± 0.01 | 0.5 ± 0.1 | $0.3^{+0.6}_{-0.15}$ |

**Notes.** The $H$ Magnitues are derived from DEEP photometry and geometry derived from best-fit orbits. The sizes are computed assuming albedos of 0.15—This is the dominant source of error in our diameter estimation.
[a] Temporary internal designation.

We compute this value with geometry obtained using the OpenOrb (M. Granvik et al. 2009) ephemerides and the apparent VR magnitudes from photometry from our DECam observations. We use these H magnitude values to estimate the diameter for each object, assuming an albedo of 0.15. These derived values are also given in Table 1. Shown in Figure 3 and overplotted on the phased DECam lightcurves are the Lowell Discovery Telescope lightcurves phased to the same period solutions. The agreement between the two data sets is excellent.

Because the measured SFR incidence rate of 0.4% is derived from a relatively small sample of 3/686 asteroids, it is most appropriate to evaluate the precision of this count using the binomial distribution. We numerically compute the integral of the binomial distribution to find the bounds of the 1 σ confidence interval, which we report as our errors (A. J. Burgasser et al. 2003). We find our probability distribution to be asymmetric due to the relatively small sample size: our final reported SFR incidence rate is $0.4^{+0.1}_{-0.3}\%$. As discussed below, future investigations of a larger sample size will considerably reduce the uncertainty of this value. We also note that this result is affected by observational biases—telescope surveys preferentially sample large objects, nearby, objects, and high-albedo objects, so these results may not be fully representative of the MBA population (R. Jedicke et al. 2016).

## 4. Strengths

Since the discovery of the first known SFR 2001 $OE_{84}$ by P. Pravec et al. (2002), the property or mechanism allowing such objects to break the spin barrier has been the subject of debate and study. If an object were monolithic in structure, perhaps a collisional fragment from a coherent parent body, then this would imply that the fragment would have much greater cohesive strength than a rubble pile. This would allow for faster rotation than the assumed critical spin rate. However, it is difficult to explain how such objects can exist in the present-day collisional environment of the main belt. The collisional lifetimes of objects of this size range are much shorter than the lifetime of the solar system. They would be highly likely to have undergone multiple disruption events in this time (G.C. de Elía & A. Brunini 2007). Although D. Polishook et al. (2017) have argued that a monolithic asteroid best explains the rotational behavior of 2001 OE$_{84}$ due to the high required strengths, the required values are still significantly less than the cohesive strength of solid rocky material (of order MPa).

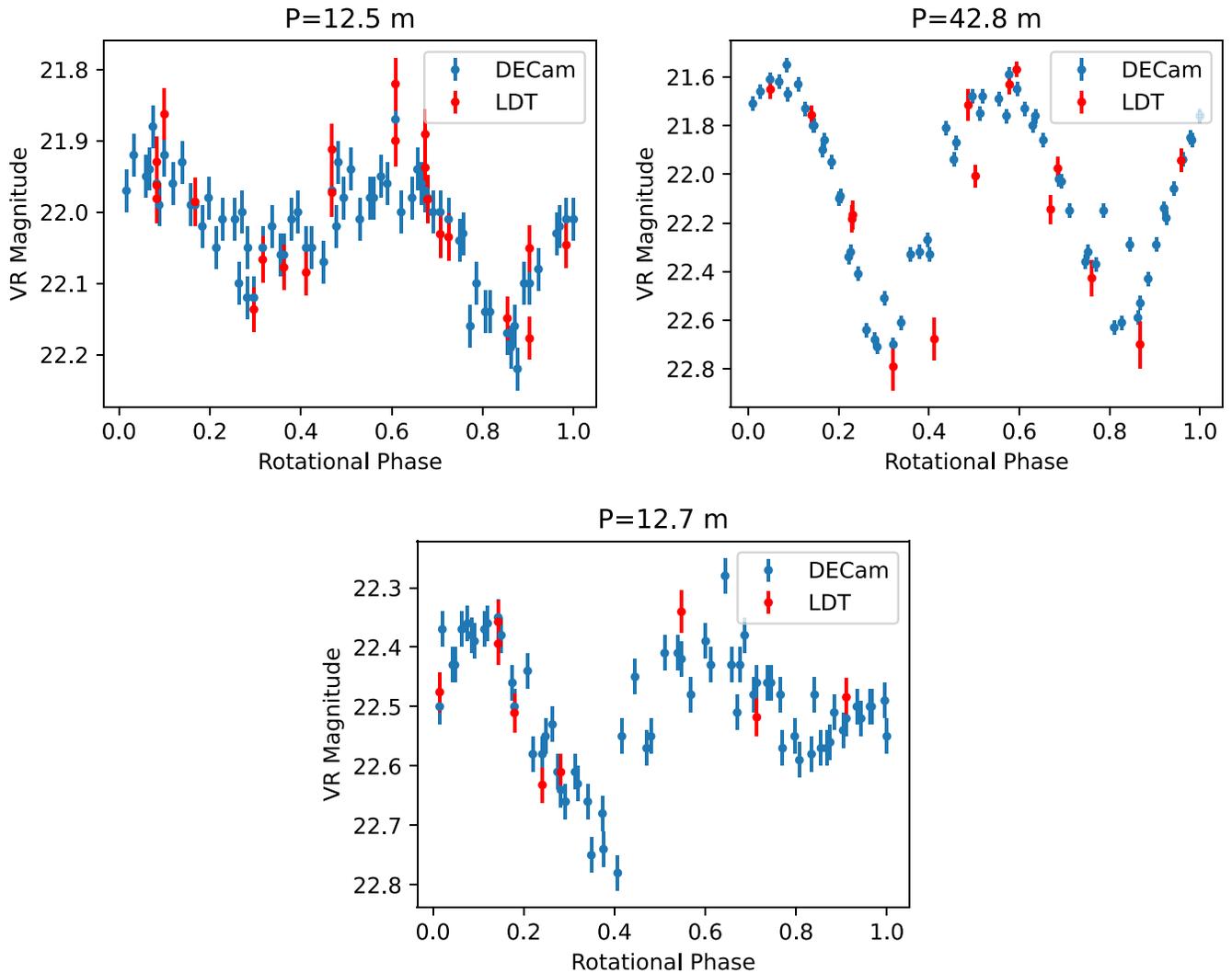

**Figure 3.** Lightcurves for the three SFRs discussed here. The lightcurves have been phased to the best-fit periodogram solutions for each rotation period. Photometry from the original DECam detection is shown in blue. Subsequent Lowell Discovery Telescope measurements, shown in red, are in excellent agreement with the DECam photometry.

Another possibility is that the SFR resists flying apart by some cohesive strength in its internal structure. This is more in line with the literature values for strength derived to date, which are of order 100–1000 s Pa. To constrain the potential cohesive strengths of these rotating ellipsoids we use a simplified form of the Drucker–Prager model, a three-dimensional model estimating the stresses within a geological material at its critical rotation state (K. A. Holsapple 2007; L. R. Alejano & A. Bobet 2012). We use a Monte Carlo numerical simulation based upon this to determine solutions for a range of values using the uncertainties in each of the asteroid parameters to constrain the required cohesion of each object. The strength values calculated represent a lower limit on the strength necessary for the object to resist rotational fission, rather than a direct measurement. The true strength of the object could be much greater than the values determined. A significant cohesive strength without a clear mechanism could indicate a monolithic structure, but only where the derived values far exceed that of regolith. A more specific discussion of the methodology used can be found in A. McNeill et al. (2018a). Figure 4 shows the minimum required strengths and confidence intervals for each of the three DEEP SFRs as a function of their density. For almost every density, a nonzero minimum cohesive strength is required for each SFR to prevent breakup.

### 5. Discussion

The possible cohesive strengths for these SFRs range from 100 to almost 10,000 Pa (Figure 4). These strengths are broadly consistent with those derived for Near-earth asteroid (162173)

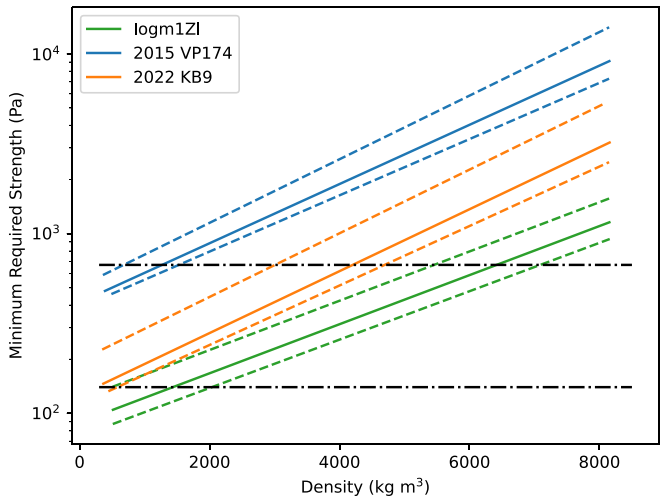

**Figure 4.** Minimum cohesive strength ranges for three DEEP SFRs as a function of bulk density. The solid lines indicate the best-fit solutions, while the dashed lines show the one-sigma uncertainty regions. The red vertical lines denote the limits of "typical" asteroid density ranges. The black horizontal lines indicate the range of likely subsurface strengths for Ryugu (M. Arakawa et al. 2020). For comparison purposes, the strength curves of all three SFRs are plotted on the same axes. Each set of three lines in matching colors corresponds to an individual asteroid.

Ryugu during the Hayabusa2 encounter (M. Arakawa et al. 2020). In that work, they found a surface regolith layer with a strength of ∼1000 Pa, with subsurface material strength closer to 500 Pa. These strength values are corroborated by laboratory studies on asteroid simulant (J. Brisset et al. 2022). While rubble piles are often considered to be completely strengthless, these results suggest that many rubble piles may have fairly significant strengths. This implies the presence of many small particles in the interior of rubble piles contributing to significant inter-particle cohesive forces. Other results imply much lower cohesive strength of regolith: deep Impact, OSIRIS-Rex, and DART all measured very low surface regolith cohesion (fewer than a few pascals) on their respective targets (K. A. Holsapple & K. R. Housen 2007; M. E. Perry et al. 2022; S. D. Raducan et al. 2024).

An important caveat is that our photometric studies only measure the bulk cohesion of a rotating body, while the laboratory and in situ measurements are extremely local and small scale. Real asteroids are extremely heterogeneous and complex bodies with various grain sizes, complicated shapes, and nonuniform compositions between individual bodies. The importance of bulk cohesion is also strongly dependent on the scale of the system: massive objects may be more dominated by gravitational forces rather than inter-particle cohesion, while the opposite is true for smaller objects (K. A. Holsapple 2007). Furthermore, simulant studies and numerical simulations demonstrate a strong grain-size dependence on cohesion, with smaller grain sizes providing significantly more cohesive strength than larger particles. P. Sánchez & D.J. Scheeres (2014) suggest that in rubble-pile asteroids with mixed grain sizes, small micron-scale grains may effectively "cement" the larger boulders together, considerably enhancing the cohesive strength. Following this, we hypothesize that the overall cohesive strength of rubble-pile asteroids may be correlated with the distribution of grain sizes within their interiors. By imposing limits on the cohesive strength of these objects, SFR period determination may be an effective indirect metric for the grain-size distribution of their interiors. The low-strength regolith measured by several spacecraft (K. A. Holsapple & K. R. Housen 2007; M. E. Perry et al. 2022; S. D. Raducan et al. 2024) suggests that cohesive strength is strongly depth dependent with a layer of recently processed regolith sitting on a more densely packed material, the depth and strength of which encodes the collisional and weathering history of that surface.

Beyond just constraining the cohesive strength, the distribution of spin periods in the DEEP data set is a probe of the overall interior density. For a given bulk density, the spin barrier can be split into two separate regimes. Above a certain cutoff size (roughly 10 km), gravity is the dominant force contributing to the strength of a rubble pile, and the minimum spin period is independent of size. Conversely, the object's overall strength below that cutoff is limited by inter-particle cohesion, and the spin barrier becomes size dependent. This is apparent in the upper envelope of the distribution in Figure 1. The value of this "spin-barrier turnoff" diameter constrains the bulk density of a given population: more dense objects will become strength-dominated at a smaller diameter than less dense objects due to their stronger gravity. Density measurements in turn provide indirect insight into the overall composition of asteroids and their taxonomies.

The overall incidence rate of SFRs in this preliminary DEEP sample also has implications for our understanding of MBAs' evolutionary processes. We identified three SFRs during a search of 686 asteroids in the DEEP data set, implying an occurrence rate of $0.4^{+0.1}_{-0.3}$%. This is greater than the overall SFR occurrence rates measured by studies such as C.-K. Chang et al. (2016) or in the LDCB itself, both of which find that less than 0.1% of rubble-pile asteroids have spin periods shorter than 2.2 hr. While this relative enhancement may in part be due to DEEP's exceptional sensitivity to SFRs, we also suggest that additional physical processes contribute to this result. The two most effective ways to spin up a main-belt asteroid are via the Yarkovsky–O'Keefe–Radzievskii–Paddack (YORP) effect (D. P. Rubincam 2000) and collisions (A. W. Harris 1979; M. Yanagisawa 2002). Both of these effects are strongly size dependent, and consequently, we expect that spin-up affects smaller subkilometer asteroids to a much greater degree than larger asteroids. The incidence rate of SFRs is strongly correlated with the size regime being investigated. DEEP has a single-exposure sensitivity of VR ∼ 23.5, which corresponds to ∼500 m in the main belt; searching for SFRs within the larger DEEP MBA sample will further constrain this result and improve our statistics, providing a more conclusive result for the number of SFRs in the main belt.

Most of LCDB SFRs Figure 1 of similar size to or smaller than the DEEP objects are NEOs. The question follows of whether NEOs are more enhanced in SFRs than MBAs, in a given size regime. If this question were answered, it could disentangle whether YORP or collisions dominate spin-up. Because NEOs orbit closer to the Sun, they are much more susceptible to thermal effects such as YORP, but MBAs are in a much more collisional regime of the solar system (D. P. Rubincam 2000; W. F. Bottke et al. 2015). If the SFR rate is similar between NEOs and MBAs, that would imply that collisions dominate and fast-rotating NEOs are first spun up in the main belt before being deposited into the near-Earth region. Conversely, if YORP is the key spin-up mechanism for creating SFRs, small MBAs should have a lower incidence rate of fast rotators than NEOs. We are well poised to address this question because of DEEP's unique sensitivity to NEO-sized MBAs. The full DEEP data set contains lightcurves for more than 60,000 main-belt objects—extrapolating the 0.4% incidence rate found here to the full data set, we expect to recover as many as 300 SFRs in the main belt. Future DEEP work will involve combining results with those of other surveys, to help determine whether the SFR rate between NEOs and MBAs differs significantly.

## 6. Conclusions and Future Work

In this work, we present our findings from an analysis of SFRs discovered in a small subset of our DEEP survey data. We find that three out of 686 asteroids examined possess rotational periods shorter than the 2.2 hr spin barrier, implying a roughly 0.4% incidence rate. Nonzero cohesive strengths are required at all physically reasonable densities for each of these three asteroids to produce the observed lightcurves. The minimum required strengths range from $\sim$100 to $\sim$10,000 Pa. While these strengths are well beyond those typical of weak lunar regolith, similar strengths have been reported from both in situ measurements and laboratory studies of regolith simulant.

The DEEP survey has a cadence and sensitivity well suited to detecting short-period variations in the lightcurves of subkilometer MBAs. The full DEEP survey has measured more than 60,000 main-belt asteroid lightcurves. Further analysis of the remaining survey data will significantly constrain both of the main results from this work: the larger sample will provide a much higher-confidence estimate of the SFR occurrence rate and will give us many more minimum required strength measurements to better characterize the interior structure of the MBA population as a whole. Assuming the SFR occurrence rate across the entire survey is consistent with the roughly 0.4% measured in this work, we expect the full DEEP survey will reveal as many as 300 SFRs in the main belt, providing us with a more robust distribution of required cohesive strengths within the main belt. The shape of this distribution for any given population will provide a probe of the rough interior structure and in turn the collisional history of objects within that population.

## Acknowledgments

This work is based in part on observations at Cerro Tololo Inter-American Observatory at NSF's NOIRLab (NOIRLab Prop. ID 2019A-0337; PI: D. Trilling), which is managed by the Association of Universities for Research in Astronomy (AURA) under a cooperative agreement with the National Science Foundation.

These results made use of the LDT at Lowell Observatory. Lowell is a private, nonprofit institution dedicated to astrophysical research and public appreciation of astronomy and operates the LDT in partnership with Boston University, the University of Maryland, the University of Toledo, Northern Arizona University and Yale University.

This work is supported by the National Aeronautics and Space Administration under grant Nos. NNX17AF21G and 80NSSC20K0670 issued through the SSO Planetary Astronomy Program and by the National Science Foundation under grant Nos. AST-2009096 and AST-2107800. This research was supported in part through computational resources and services provided by Advanced Research Computing at the University of Michigan, Ann Arbor. This work used the Extreme Science and Engineering Discovery Environment (XSEDE; J. Towns et al. 2014), which is supported by National Science Foundation grant No. ACI-1548562. This work used the XSEDE Bridges GPU and Bridges-2 GPU-AI at the Pittsburgh Supercomputing Center through allocation TG-AST200009.

K.J.N. is supported by the Eric and Wendy Schmidt AI in Science Postdoctoral Fellowship, a Schmidt Futures program.

H.S. acknowledges support by NASA under grant No. 80NSSC21K1528 (FINESST). H.S., M.J., P.B., and C.C. acknowledge the support from the University of Washington College of Arts and Sciences, Department of Astronomy, and the DiRAC Institute. The DiRAC Institute is supported through generous gifts from the Charles and Lisa Simonyi Fund for Arts and Sciences and the Washington Research Foundation. M.J. wishes to acknowledge the support of the Washington Research Foundation Data Science Term Chair fund, and the University of Washington Provost's Initiative in Data-Intensive Discovery.

This project used data obtained with the DECam, which was constructed by the Dark Energy Survey (DES) collaboration. Funding for the DES Projects has been provided by the US Department of Energy, the US National Science Foundation, the Ministry of Science and Education of Spain, the Science and Technology Facilities Council of the United Kingdom, the Higher Education Funding Council for England, the National Center for Supercomputing Applications at the University of Illinois at Urbana-Champaign, the Kavli Institute for Cosmological Physics at the University of Chicago, Center for Cosmology and Astro-Particle Physics at the Ohio State University, the Mitchell Institute for Fundamental Physics and Astronomy at Texas A&M University, Financiadora de Estudos e Projetos, Fundação Carlos Chagas Filho de Amparo á Pesquisa do Estado do Rio de Janeiro, Conselho Nacional de Desenvolvimento Científico e Tecnológico and the Ministério da Ciência, Tecnologia e Inovação, the Deutsche Forschungsgemeinschaft, and the Collaborating Institutions in the Dark Energy Survey.

The Collaborating Institutions are Argonne National Laboratory, the University of California at Santa Cruz, the University of California at Los Angeles, the University of Cambridge, Centro de Investigaciones Enérgeticas, Medioambientales y Tecnológicas-Madrid, the University of Chicago, University College London, the DES-Brazil Consortium, the University of Edinburgh, the Eidgenössische Technische Hochschule (ETH) Zürich, Fermi National Accelerator Laboratory, the University of Illinois at Urbana-Champaign, the Institut de Ciències de l'Espai (IEEC/CSIC), the Institut de Física d'Altes Energies, Lawrence Berkeley National Laboratory, the Ludwig-Maximilians Universität München and the associated Excellence Cluster Universe, the University of Michigan, NSF's NOIRLab, the University of Nottingham, the Ohio State University, the OzDES Membership Consortium, the University of Pennsylvania, the University of Portsmouth, SLAC National Accelerator Laboratory, Stanford University, the University of Sussex, and Texas A&M University.

C.F. acknowledges support from the BASAL Centro de Astrofísica y Tecnologías Afines (CATA) ANID BASAL project FB210003 and from the Ministry of Economy, Development, and Tourism's Millennium Science Initiative through grant IC120009, awarded to The Millennium Institute of Astrophysics (MAS).

This paper benefited from useful feedback from an anonymous referee. We thank them for their time and thoughtful comments.

### ORCID iDs

Ryder Strauss ● https://orcid.org/0000-0001-6350-807X
Andrew McNeill ● https://orcid.org/0009-0005-9955-1500
David E. Trilling ● https://orcid.org/0000-0003-4580-3790
Francisco Valdes ● https://orcid.org/0000-0001-5567-1301
Pedro H. Bernardinelli ● https://orcid.org/0000-0003-0743-9422
Cesar Fuentes ● https://orcid.org/0000-0002-5211-0020
David W. Gerdes ● https://orcid.org/0000-0001-6942-2736
Matthew J. Holman ● https://orcid.org/0000-0002-1139-4880
Mario Jurić ● https://orcid.org/0000-0003-1996-9252
Hsing Wen Lin (林省文) ● https://orcid.org/0000-0001-7737-6784
Larissa Markwardt ● https://orcid.org/0000-0002-2486-1118
Michael Mommert ● https://orcid.org/0000-0002-7817-3388
Kevin J. Napier ● https://orcid.org/0000-0003-4827-5049
William J. Oldroyd ● https://orcid.org/0000-0001-5750-4953
Matthew J. Payne ● https://orcid.org/0000-0001-5133-6303
Andrew S. Rivkin ● https://orcid.org/0000-0002-9939-9976
Hilke E. Schlichting ● https://orcid.org/0000-0002-0298-8089
Scott S. Sheppard ● https://orcid.org/0000-0003-3145-8682
Hayden Smotherman ● https://orcid.org/0000-0002-7895-4344
Chadwick A. Trujillo ● https://orcid.org/0000-0001-9859-0894
Fred C. Adams ● https://orcid.org/0000-0002-8167-1767
Colin Orion Chandler ● https://orcid.org/0000-0001-7335-1715

### References

Alejano, L. R., & Bobet, A. 2012, RMRE, 45, 995
Arakawa, M., Saiki, T., Wada, K., et al. 2020, Sci, 368, 67
Bellm, E. C., Kulkarni, S. R., Graham, M. J., et al. 2019, PASP, 131, 018002
Bernardinelli, P. H., Smotherman, H., Langford, Z., et al. 2024, AJ, 167, 134
Bottke, W. F., Brož, M., O'Brien, D. P., et al. 2015, Asteroids IV (Tucson, AZ: Univ. Arizona Press), 701
Bradley, L., Sipőcz, B., Robitaille, T., et al. 2024, astropy/photutils: v1.11.0, Zenodo, doi:10.5281/zenodo.10671725
Brisset, J., Cox, C., Metzger, J., et al. 2022, PSJ, 3, 176
Burgasser, A. J., Kirkpatrick, J. D., Reid, I. N., et al. 2003, ApJ, 586, 512
Carry, B. 2012, P&SS, 73, 98
Chang, C.-K., Lin, H.-W., Ip, W.-H., et al. 2016, ApJS, 227, 20
de Elía, G. C., & Brunini, A. 2007, A&A, 466, 1159
DeMeo, F. E., & Carry, B. 2013, Icar, 226, 723
Durech, J., Kaasalainen, M., Warner, B. D., et al. 2009, A&A, 493, 291
Duxbury, T. C., Newburn, R. L., Acton, C. H., et al. 2004, JGRE, 109, E02002
Flaugher, B., Diehl, H. T., Honscheid, K., et al. 2015, AJ, 150, 150
Gaia Collaboration, Vallenari, A., Brown, A. G. A., et al. 2023, A&A, 674, A1
Granvik, M., Virtanen, J., Oszkiewicz, D., & Muinonen, K. 2009, M&PS, 44, 1853
Hanuš, J., Ďurech, J., Brož, M., et al. 2011, A&A, 530, A134
Harris, A. W. 1979, Icar, 40, 145
Heinze, A. N., Tonry, J. L., Denneau, L., et al. 2018, AJ, 156, 241
Holsapple, K. A. 2007, Icar, 187, 500
Holsapple, K. A., & Housen, K. R. 2007, Icar, 187, 345
Ivezić, Ž., Tabachnik, S., Rafikov, R., et al. 2001, AJ, 122, 2749
Jedicke, R., Bolin, B., Granvik, M., & Beshore, E. 2016, Icar, 266, 173
Kramer, D., Gowanlock, M., Trilling, D., McNeill, A., & Erasmus, N. 2023, A&C, 44, 100711
Law, N. M., Fors, O., Ratzloff, J., et al. 2015, PASP, 127, 234
Licandro, J., Popescu, M., Tatsumi, E., et al. 2023, MNRAS, 521, 3784
Lomb, N. R. 1976, Ap&SS, 39, 447
McNeill, A., Trilling, D. E., & Mommert, M. 2018a, ApJL, 857, L1
McNeill, A., Fitzsimmons, A., Jedicke, R., et al. 2018b, AJ, 156, 282
Monteiro, F., Silva, J. S., Tamayo, F., Rodrigues, T., & Lazzaro, D. 2020, MNRAS, 495, 3990
Napier, K. J., Lin, H. W., Gerdes, D. W., et al. 2024, PSJ, 5, 50
Perry, M. E., Barnouin, O. S., Daly, R. T., et al. 2022, NatGe, 15, 447
Polishook, D., Jacobson, S. A., Morbidelli, A., & Aharonson, O. 2017, NatAs, 1, 0179
Pravec, P., Kušnirák, P., Šarounová, L., et al. 2002, in Asteroids, Comets, and Meteors: ACM 2002, ed. B. Warmbein (Noordwijk: ESA), 743
Raducan, S. D., Jutzi, M., Cheng, A. F., et al. 2024, NatAs, 8, 445
Rondón, E., Lazzaro, D., Rodrigues, T., et al. 2020, PASP, 132, 065001
Rubincam, D. P. 2000, Icar, 148, 2
Sánchez, P., & Scheeres, D. J. 2014, M&PS, 49, 788
Scargle, J. D. 1989, ApJ, 343, 874
Sergeyev, A. V., & Carry, B. 2021, A&A, 652, A59
Smotherman, H., Bernardinelli, P. H., Portillo, S. K. N., et al. 2024, AJ, 167, 136
Strauss, R., Trilling, D. E., Bernardinelli, P. H., et al. 2024, AJ, 167, 135
Taylor, M., Altobelli, N., Martin, P., Buratti, B. J., & Choukroun, M. 2016, DPS, 48, 01
Thomas, P. C., Joseph, J., Carcich, B., et al. 2000, Icar, 145, 348
Towns, J., Cockerill, T., Dahan, M., et al. 2014, CSE, 16, 62
Trilling, D. E., Gowanlock, M., Kramer, D., et al. 2023, AJ, 165, 111
Trilling, D. E., Gerdes, D. W., Juric`, M., et al. 2024, AJ, 167, 132
Trujillo, C. A., Fuentes, C., Gerdes, D. W., et al. 2024, AJ, 167, 133
Valdes, F., Gruendl, R. & DES Project 2014, in ASP Conf. Ser. 485, Astronomical Data Analysis Software and Systems XXIII, ed. N. Manset & P. Forshay (San Francisco, CA: ASP), 379
VanderPlas, J. T. 2018, ApJS, 236, 16
Veverka, J., Thomas, P. C., Helfenstein, P., et al. 1996, Icar, 120, 200
Walsh, K. J. 2018, ARA&A, 56, 593
Warner, B. D., Harris, A. W., & Pravec, P. 2009, Icar, 202, 134
Yanagisawa, M. 2002, Icar, 159, 300